\documentstyle[preprint,aps]{revtex}

\def\yazi{\mathop{\to}}
\newcommand{\ya}{\displaystyle\yazi_{r\to\infty}}

\newcommand{\br}{\mbox{$\boldmath r$}}
\begin{document}

\draft

\preprint{}

\title{Unitarity of the Aharonov-Bohm Scattering Amplitudes} 

\author{Masato Arai and Hisakazu Minakata}
\address{Department of Physics, Tokyo Metropolitan University \\
Minami-Osawa, Hachioji, Tokyo 192-03, Japan}

\date{May, 1997}


\maketitle

\begin{abstract}
We discuss the unitarity relation of the Aharonov-Bohm scattering 
amplitude with the hope that it distinguishes between the differing 
treatments which employ different incident waves. We find that the 
original Aharonov-Bohm scattering amplitude satisfies the 
unitarity relation under the regularization prescription whose 
theoretical foundation does not appear to be understood. On the 
other hand, the amplitude obtained by Ruijsenaars who uses plane 
wave as incident wave also satisfies the unitarity relation but 
in an unusual way.   

\end{abstract}

 
\section{Introduction and Summary}

In their pioneering work Aharonov and Bohm \cite{ab} examined the 
scattering of nonrelativistic charged particles off a magnetic flux 
of infinitesimal radius. We shall call the process as the AB scattering 
in this paper. At the very least the problem serves as an idealized 
system which exhibits the Aharonov-Bohm effect. 

While more than one-third of the century has been passed since their 
paper there seem to still exist some disagreements among the literatures 
on the treatment and the interpretation of the AB scattering. 
Ruijsenaars \cite{snm} and others\cite{so} advocates the viewpoint that 
one has to take plane wave as incident wave, as opposed to the original 
treatment by Aharonov and Bohm. These authors' treatment entails the 
S-matrix which contains a delta-function peaked in the forward direction. 

The treatment of the AB scattering with incident plane wave was 
critically examined by Hagen\cite{hagen}. He pointed out that taking 
the asymptotic limit $r \rightarrow \infty$ and summing over angular 
momenta do not commute with each other owing to the violent infrared 
(= high angular momentum) behavior of the scattering amplitude. 
Thus, the usual definition of the phase shift, which involves the 
procedure of taking asymptotic limit in each partial wave, does not 
work. The observation casts serious doubt on the treatment with 
incident plane wave, but it does not appear to be the last ward to 
settle the controversy. 

In this paper we examine the unitarity of the the S-matrix of the AB 
scattering in hoping that it may discriminate differing treatments in 
the literatures. There is a ``naive'' argument that the delta-function 
in the forward direction is physically meaningless because it cannot 
be directly observed. This is not correct because the forward scattering 
amplitude is related with the total cross section by the optical theorem. 
This was the original motivation which leads us to the study of unitarity 
relation of the AB scattering amplitude. 

In fact, the problem is slightly more complicated. As some readers 
might have noticed the original AB scattering amplitude diverges in 
the forward direction, rendering the detection of the delta-function 
contribution difficult. However, it is also true that the unitarity 
relation in non-forward direction contains the information of forward 
scattering amplitude in its right-hand-side (RHS). (See below.) 
Therefore, it appears that the setting of the problem itself seems 
to be meaningful and our investigation has started along this line 
of thought. 

In carrying out this investigation we have encountered the new 
feature of the problem that we never expected before actually 
engaging the work. Our conclusions at hand are as follows:

\noindent
(1) With the choice of plane wave as incident wave the unitarity 
relation of the AB scattering amplitude holds but in a contrived 
way as will be explained in Sec. V. 

\noindent 
(2) The scattering amplitude obtained by Aharonov and Bohm satisfies 
the unitarity relation if one employs the suitable regularization 
prescription which is consistent with positivity of RHS of the 
unitarity relation near the forward direction. Unfortunately, we 
fail in our attempts at deriving it in a physically reasonable
way and thereby placing the regularization prescription on a firm 
theoretical ground, as we will describe in Secs. III and IV. 

Thus, we have not made our original goal of distinguishing between 
two different choices of incident wave. Instead, we learn some 
lessons, and in particular uncover the necessity of an 
``i$\epsilon$-prescription'' which, to our knowledge, does not seem 
to be noticed before.

In Sec. II the basic formulas of the AB scattering problem are 
briefly reviewed to define our notations. We follow the notation 
of the original paper by Aharonov and Bohm \cite{ab}. 
The expressions of scattering amplitudes by Aharonov and Bohm 
and by Ruijsenaars are also recollected. In Sec. III 
the unitarity relation of Aharonov-Bohm's scattering amplitude is 
examined. The necessity of a phenomenological regularization 
prescription is noted. In Sec. IV some unsuccessful trials for 
deriving the regularization prescription by exploiting hard-core 
potential, or finite-radius magnetic flux are described. In Sec. V 
the unitarity relation of Ruijsenaars' scattering amplitude is 
discussed. Sec. VI summarizes our investigations.  
In Appendix the unitarity relation of the AB scattering amplitude 
with modified incident wave is derived by utilizing the method of  
Landau and Lifshitz. 

\section{The Aharonov-Bohm Scattering Amplitudes}
 
We follow the original notation of Aharonov and Bohm and denote 
the flux parameter as $\alpha = - e\Phi /2\pi$ 
where $\Phi$ is the magnetic flux. 
The Schr\"odinger equation takes the form in polar coordinate as  
\begin{equation}
\displaystyle\left[\frac{1}{r}\frac{\partial}{\partial r}
\left(r\frac{\partial}{\partial r} \right)+\frac{1}{r^2}
\left(\frac{\partial}{\partial \theta}+i\alpha \right)^2+k^2 \right]
\psi=0, 
\label{eq:scheq}
\end{equation}
where we treat the problem in a two-dimensional setting by ignoring 
the separated z-direction and $k$ is the wave number. The regular 
solution of the Schr\"odinger equation which vanishes at the origin 
(i.e., the location of the magnetic flux) can be written as
\begin{equation}
\psi_{AB}(r,\theta)=\displaystyle\sum_{m=-\infty}^{\infty}(-i)^{|m+\alpha|}
J_{|m+\alpha|}(kr)e^{im\theta} \label{eq:regular}
\label{eq:psi}
\end{equation}

There is a variety of ways of extracting the asymptotic form of the wave 
function at spatial infinity; the original method of Aharonov and Bohm 
\cite{ab}, the method of contour deformation by Berry and coworkers 
\cite{bcl}, and the Takabayasi method\cite{taka} which utilizes an 
integral representation of the Bessel function. 
All these methods agree with each other and result in the expression 
\begin{equation}
\psi_{AB}(r,\theta)\ya e^{-i(kr\cos\theta+\alpha\theta)}+
\displaystyle\frac{e^{ikr}}{\sqrt{r}}f_{AB}(\theta)
\label{eq:scattering1}
\end{equation}
\begin{equation}
f_{AB}(\theta)=\frac{-1}{\sqrt{2\pi k}}e^{\frac{i}{4}\pi}(-1)^{-[\alpha]}
e^{-i([\alpha]+\frac{1}{2})\theta}\frac{\sin\pi\alpha}{\cos\frac{\theta}{2}}
\label{eq:scattering2}
\end{equation}
where $[\alpha]$ denotes the largest integer which is less than or equal 
to $\alpha$. Notice that we take the convention that the incident wave 
moves to the negative $x$ direction and the forward scattering corresponds 
to $\theta=\pm\pi$. 

On the other hand, the asymptotic form obtained by Ruijsenaars \cite{snm} 
takes the form 
\begin{equation}
\psi_{AB}(r,\theta)\ya e^{-ikr\cos\theta}+ 
\displaystyle\frac{e^{ikr}}{\sqrt{r}}f_{R}(\theta) \label{eq:asy}
\end{equation}
\begin{eqnarray}
f_R(\theta)&=&-\displaystyle\sqrt{\frac{2\pi}{k}}
e^{-\frac{i}{4}\pi}
{\Bigg[} 
(1-\cos\pi\alpha)\delta(\theta -\pi)\nonumber\\
&&\hskip 2cm + \
\frac{i}{\pi}(-1)^{[\alpha]}
\sin\pi\alpha e^{-i[\alpha]\theta}
P[\frac{1}{e^{i\theta}+1}]\;\;
{\Bigg]}
\label{eq:delta}
\end{eqnarray}
where $P$ denotes the principal value prescription. 

An important distinction between the Aharonov-Bohm and the Ruijsenaars 
scattering amplitudes is that the former is not defined at the forward 
direction whereas the latter is. It is implicit \cite{error} in the 
original paper by Aharonov and Bohm \cite{ab} and was emphasized in 
\cite {bcl} that the scattering amplitude (\ref{eq:scattering1}) 
is defined except for the narrow cone $|\theta|>\pi - O[(kr)^{-1/2}]$. 
While it is defined in the forward direction, the square modulus of 
the Ruijsenaars scattering amplitude does not appear to be well defined. 
A possible way of obtaining finite forward scattering amplitude by 
modifying the boundary condition at the origin has recently been put 
forward by Giacconi et al. \cite {GMS}. It is also argued by Stelitano 
\cite {Stelitano} that the time-dependent formulation is necessary 
for consistent treatment of the AB scattering at around the forward 
direction.

\section{Unitarity of the Aharanov-Bohm Scattering Amplitude;
Phenomenological Approach}

Let us define the scattering amplitude (\ref{eq:scattering1}) apart 
from the forward direction, or more precisely, up to an infinitesimal 
value of $\pi - \theta$, by defining it at $r \rightarrow \infty$. 
We then discuss the unitarity relation of the Aharonov-Bohm scattering 
amplitude. 

In Appendix we follow the method described by Landau and Lifshitz 
\cite{LL} to derive the unitarity relation of the scattering amplitude 
corresponding to the choice of incident wave as in (\ref{eq:scattering1}). 
It reads
\begin{equation}
\label{eq:scattering3}
e^{i(\alpha-\frac{1}{4})\pi}f(\theta)-
e^{-i(\alpha-\frac{1}{4})\pi}f^*(-\theta)=
i\sqrt{\frac{k}{2\pi}}\int^{\pi}_{-\pi}d\theta'
f^*(\theta')f(\theta'+\theta+\pi)
\end{equation}

We have to remark that we take the viewpoint in this paper that 
the scattering amplitude in (\ref{eq:scattering1})

Using (\ref{eq:scattering2}) the left-hand-side (LHS) of the unitarity 
relation can be expressed as
\begin{equation}
\mbox{LHS}=-i\sqrt{\frac{2}{\pi k}}(-1)^{[\alpha]}
e^{{-i}([\alpha]+\frac{1}{2})\theta}
\frac{\sin^2\pi\alpha}{\cos\frac{\theta}{2}}.
\label{eq:lhs}
\end{equation}
To compute RHS the expression of the scattering amplitude 
(\ref{eq:scattering2}) is not enough; we need to specify certain 
``i$\epsilon$ prescription'' to dictate how to make detour around 
the singularities. Notice that the Ruijsenaars amplitude does have 
such prescription, taking the principal value, as indicated in 
(\ref{eq:delta}). 

Lacking any known ``i$\epsilon$-prescriptions'' we try to identify it via 
the Berry et al.'s method for deriving the asymptotic form of the wave 
function. They use the integral representation of the Bessel function 
\cite{Bateman}
\begin{equation}
\label{eq:asympt}
J_\nu(z)=\frac{1}{2\pi}\int_C dt \exp[i(\nu t-z\sin t)]
\end{equation}
where $C$ is the contour starting from $-\pi+i\infty$ and goes down 
to $-\pi$ and traverse to $+\pi$ on the real axis, and then goes up to 
$+\pi+i\infty$. If we insert (\ref{eq:asympt}) into (\ref{eq:regular}) 
the summation over m converges if we add small positive imaginary part 
on the contour along the real axis. One obtains 
\begin{eqnarray}
\psi_{AB}&=&\frac{1}{2\pi}\int_C dt e^{-ikr\sin t}
\left[
\frac{\exp\{-i[(t-\frac{\pi}{2})(\alpha-[\alpha]-1)+([\alpha]+1)\theta]\}}
{1-\exp[i(t-\frac{\pi}{2}-\theta)]}\right.\nonumber\\
&&\hskip 5cm+ \left.
\frac{\exp\{i[(t-\frac{\pi}{2})(\alpha-[\alpha])-[\alpha]\theta]\}}
{1-\exp[i(t-\frac{\pi}{2}+\theta)]}\;\;
\right]
\label{eq:psiAB}
\end{eqnarray}
Note that we differ in sign from \cite{bcl} in defining the flux 
parameter $\alpha$. 

Then, we deform the contour $C$ into $C'$ which passes through 
$-\frac{\pi}{2}+i\epsilon$ and moves down into the lower $t$-plane 
and again goes up to the top of the upper $t$-plane by passing 
through $+\frac{\pi}{2}+i\epsilon$, as described in \cite{bcl}. 
Through the process of the deformation we pick up a 
pole at somewhere on the real axis $-\frac{\pi}{2}<t<\frac{\pi}{2}$. 
The pole term comes from the first (second) term in (\ref{eq:psiAB}) 
provided that $-\pi<\theta<0 (0<\theta<\pi)$. As shown by Berry et al. 
the pole term gives rise to the incident wave of Aharonov and Bohm 
as in (\ref{eq:scattering1}). The remaining contribution comes from 
the saddle point at $t=\pm\frac{\pi}{2}$. The saddle point at 
$t=-\frac{\pi}{2}$ produces the scattering wave (\ref{eq:scattering2}). 
The saddle point at $t=\frac{\pi}{2}$, which is potentially dangerous 
because it would produce incoming wave, makes no contribution owing to 
the vanishing residue. We believe it natural to keep small positive 
imaginary part in computing the saddle-point contribution at 
$t=-\frac{\pi}{2}$. Namely, we do saddle-point integration at 
$t=-\frac{\pi}{2}+i\epsilon$. It lead to the regularized form of 
the AB scattering amplitude 
\begin{eqnarray}
f_{AB}(\theta) &=& -\frac{1}{\sqrt{2\pi k}}
e^{-\frac{i}{4}\pi}(-1)^{[\alpha]}
e^{-i[\alpha]\theta}\nonumber\\
&&\hskip 1cm\times 
\left[ \frac{1}{1+e^{i(\theta-i\epsilon)}}e^{i\alpha\pi}
-\frac{1}{1+e^{i(\theta+i\epsilon)}}e^{-i\alpha\pi}\right]
\label{eq:ab}
\end{eqnarray}

Having specified the regularization prescription we are ready to 
compute RHS of the unitarity relation. Alas we have a trouble; 
RHS vanishes. 

Since the simplest regularization prescription fails we look for 
a phenomenologically successful regularization prescription. 
It turns out that the solution is given by the $\theta-i\epsilon$ 
prescription. That is, we replace $e^{i(\theta+i\epsilon)}$ in the 
second term in the square bracket in (\ref{eq:ab}) into 
$e^{i(\theta-i\epsilon)}$. (Note that the first term already meets 
the requirement.) Under the regularization procedure just specified 
one can easily compute RHS of the unitarity relation. 
Changing the integration variable into $z=e^{i\theta'}$ it can be 
expressed as
\begin{equation}
\label{eq:rhs}
\mbox{RHS}=\frac{-4}{\sqrt{(2\pi)^3 k}}(-1)^{\alpha}e^{-i[\alpha]\theta}
\sin^2\pi\alpha
\oint dz \frac{e^{-i\theta}}{(z+e^\epsilon)(z-e^{-i\theta}e^{-\epsilon})}
\end{equation}
where the integration contour is along the circle $|z|=1$. 
The $i\epsilon-$prescription dictates to pick up the pole at 
$z=e^{-i\theta}$ and the resulting expression of RHS coincides 
with LHS in (\ref{eq:lhs}). Thus, we have shown that the Aharonov-Bohm 
scattering amplitude satisfies the unitarity relation provided that 
the phenomenological $\theta-i\epsilon$ prescription is employed.

\section{Looking for Regularization Prescription}

It is natural to expect that the $\theta - i\epsilon$ prescription 
can naturally be derived from certain physical regularization 
procedures which are able to regulates the divergence of forward 
scattering amplitude. The most natural possibility is to introduce 
a small but finite radius of the magnetic flux. Unfortunately, 
the solution becomes complicated. Therefore, we postpone the 
investigation of this case to the end of this section and start 
with the simpler problem of setting up the hard core potential of 
radius $R$, keeping the width of the magnetic flux infinitesimal. 
The exact solution of this problem is again given by Berry et al 
\cite{bcl}. The wave function takes the form
\begin{eqnarray}
\psi(\br) &=& \psi_{AB}(\br)-\psi_R(\br)\\
\psi_R(\br) &=& \sum_{m=-\infty}^{+\infty}
b_m (-i)^{|m+\alpha|}e^{im\theta}H^{(1)}_{|m+\alpha|}(kr)
\label{eq:psiR1}
\end{eqnarray}
where 
\begin{equation}
b_m = \frac{J_{|m + \alpha|}(kR)}{H_{|m + \alpha|}^{(1)}(kR)}.
\end{equation}
Here, $H^{(1)}(z)$ is the Hankel function of the first kind. 

We use the integral representation of the Hankel function \cite{Bateman} 
\begin{equation}
H_\nu^{(1)}(z) = \frac{1}{\pi}\int_{C_1}dt \exp[i(\nu t-z\sin t)].
\end{equation}
The contour $C_1$ runs from $t=x_{+} + i\infty$ to $t=x_{-}-i\infty$ by 
passing through $t=-\frac{\pi}{2}$, where $-\pi<x_{+}<-\frac{\pi}{2}$ 
and $-\frac{\pi}{2} < x_{-}< 0$. We compute $\psi_R$ by using the 
saddle-point approximation at $t=-\frac{\pi}{2}$ to evaluate the 
nonzero $R$ correction to the AB scattering wave. 
For this purpose it suffices to keep the leading order in $kR$ in 
(\ref{eq:psiR1}). Then, the integrand does not develop pole 
singularities unlike the case of evaluating $\psi_{AB}$. The dominant 
contribution comes from the saddle point at $t=-\frac{\pi}{2}$. 
Using the small $z$ behavior of the Bessel functions 
\begin{equation}
\frac{J_{\nu}(z)}{H_{\nu}^{(1)}(z)} \;
\displaystyle\yazi_{z\to\ 0}\; 
\frac{i\pi}{\Gamma(\nu)\Gamma(1+\nu)}
\left(\frac{z}{2}\right)^{2\nu}
\end{equation}
we obtain 
\begin{eqnarray}
\psi_R(\br) && 
\ya i e^{ikr} \left(\frac {2 \pi}{ikr}\right)^\frac {1}{2}
\displaystyle\left[ 
\frac{e^{-i[\alpha]\theta}e^{-i\pi(\alpha-[\alpha])}}
{\Gamma(\alpha-[\alpha])\Gamma(\alpha-[\alpha]+1)}
\left(\frac{kR}{2}\right)^{2(\alpha-[\alpha])}\right .\nonumber\\
&& \hspace{2cm}\left .-
\frac{e^{-i([\alpha]+1)\theta}e^{i\pi(\alpha-[\alpha])}}
{\Gamma([\alpha]-\alpha+1)\Gamma([\alpha]-\alpha+2)}
\left(\frac{kR}{2}\right)^{2([\alpha]-\alpha+1)}\right]
\label{eq:psiR2}
\end{eqnarray}

On the other hand the order $\epsilon$ correction expected from 
the $\theta-i\epsilon$ prescription takes the form 
\begin{equation}
\psi(\br)\ya\psi_{AB}(\br)- 
\left(\frac {2 \pi}{ikr}\right)^\frac {1}{2}
\frac{e^{ikr}}{8 \pi \cos^2(\frac{\theta}{2})}\left[
\epsilon e^{-i([\alpha]]+1)\theta} e^{i\pi(\alpha-[\alpha])}
+ \delta e^{-i[\alpha]\theta}e^{-i\pi(\alpha-[\alpha])}\right]
\label{eq:psireg}
\end{equation}
where we have regulated the first and the second terms of 
(\ref{eq:psiAB}) by replacing $\theta$ by $\theta-i\epsilon$ and 
$\theta-i\delta (\epsilon > 0, \delta > 0)$, respectively. 

In spite of their similarity, the correction terms in (\ref{eq:psiR2}) 
and (\ref{eq:psireg}) differs by two important respects; 
The two terms differ in sign in (\ref{eq:psiR2}) whereas those in 
(\ref{eq:psireg}) have the same relative sign. Also there exists an 
extra over-all $i$ in (\ref{eq:psiR2}) relative to the correction 
terms in (\ref{eq:psireg}). 
Thus, we conclude that the finite width regularization does not give 
rise to the requested $\theta-i\epsilon$ prescription. 

One can repeat the similar calculation for the case of finite radius 
magnetic flux. In this case the vector potential may be taken as
\begin{equation}
A_{\theta}(r)=
\left\{
\begin{array}{ll}
\displaystyle\frac{\Phi r}{2\pi R^2} & (r<R)\\
\displaystyle\frac{\Phi}{2\pi r} & (r>R)
\end{array}
\right.
\end{equation}
The solution of the Schr\"odinger equation in the outer region is 
given by (\ref{eq:psiR1}). 
The one in the inner region is given by the Whittaker function 
\begin{equation}
\psi_{inner}(\br)=
\frac{1}{\sqrt{\alpha}}\left(\frac{R}{r}\right)\sum_{m=-\infty}^{+\infty}a_m
(-i)^{|m+\alpha|}e^{im\theta}
M_{\lambda-\frac{m}{2},\frac{|m|}{2}}\left[\alpha\left(\frac{r}{R}\right)^2\right]
\end{equation}
where $\lambda=(kR)^2/4\alpha$. 
One can determine $a_m$ and $b_m$ by matching the wave functions and their 
derivatives at $r=R$. 
We obtain 
\begin{equation}
b_m=\frac
{J_{|m+\alpha|}(kR)\left\{2\alpha M^\prime_{\lambda-\frac{m}{2},\frac{|m|}{2}}(\
\alpha)
-M_{\lambda-\frac{m}{2},\frac{|m|}{2}}(\alpha)\right\}
-kRJ^\prime_{|m+\alpha|}(kR)M_{\lambda-\frac{m}{2},\frac{|m|}{2}}}
{H_{|m+\alpha|}(kR)\left\{2\alpha M^\prime_{\lambda-\frac{m}{2},\frac{|m|}{2}}(\
\alpha)
-M_{\lambda-\frac{m}{2},\frac{|m|}{2}}(\alpha)\right\}
-kRH^{(1)\prime}_{|m+\alpha|}(kR)M_{\lambda-\frac{m}{2},\frac{|m|}{2}}}
\end{equation}
and a similar expression for $a_m$. 

We can go through the same analysis as before and we end up the same 
result as in (\ref{eq:psiR2}) but with the first and the second terms 
in (\ref{eq:psiR2}) being multiplied by
\begin{equation}
\frac{([\alpha]+|[\alpha]|+1) M_{\kappa+1,|\kappa|}(\alpha) 
- M_{\kappa,|\kappa|}(\alpha)}
{(|[\alpha]|-[\alpha]+1)M_{\kappa-1,|\kappa|}(\alpha) 
- M_{\kappa,|\kappa|}(\alpha)}
\end{equation}
and
\begin{equation}
\frac
{(|[\alpha]+1|-[\alpha])M_{\kappa-\frac{1}{2},|\kappa+\frac{1}{2}|}(\alpha)
- M_{\kappa+\frac{1}{2},|\kappa+\frac{1}{2}|}(\alpha)}
{(|[\alpha]+1|+[\alpha]+2)M_{\kappa+\frac{3}{2},|\kappa+\frac{1}{2}|}(\alpha)
- M_{\kappa+\frac{1}{2},|\kappa+\frac{1}{2}|}(\alpha)},
\end{equation}
respectively, where $\kappa\equiv \displaystyle\frac{[\alpha]}{2}$.
Because of the surviving ``$i$-problem'' the finite width magnetic flux 
does not give the required $\theta-i\epsilon$ prescription.

\section{Unitarity of the Ruijsenaars Amplitude}

The unitarity relation of the Ruijsenaars amplitude is even more 
subtle. Having employed the plane wave as incident wave the unitarity 
relation is different from (\ref{eq:scattering3}); one can repeat the 
same procedure as before and the result is obtained by setting 
$\alpha=0$ in (\ref{eq:scattering3}). Then, we have a curious result. 
LHS vanishes except for the forward direction, 
\begin{equation}
\label{eq:lhs2}
\mbox{LHS}=2i\sqrt{\frac{2\pi}{k}}(1-\cos\pi\alpha)\delta(\theta-\pi).
\end{equation}
One can easily show that RHS gives rise to the same expression as 
(\ref{eq:lhs2}) thanks to the principal value prescription. Therefore, 
the unitarity relation holds in the sense that both LHS and RHS give 
the identical delta-function contribution in the forward direction, 
and vanish elsewhere. We do not know any other examples of scattering 
problem whose scattering amplitude possesses such curious property. 

\section{Conclusion}

We summarize our investigation in this paper.

\noindent
(1)\quad The scattering amplitude obtained by Aharonov and Bohm 
satisfies the unitarity relation under the phenomenological 
$\theta-i\epsilon$ prescription. 
Unfortunately, we neither succeeded to systematically derive the 
prescription nor pinned down its physical meaning. 

\noindent
(2)\quad The Ruijsenaars scattering amplitude obeys unitarity 
relation in a contrived way that both LHS and RHS vanish anywhere 
at $\theta\neq\pi$. 

\noindent
(3)\quad To our understanding of the problem the unitarity relation 
of the AB scattering amplitude does not appear to select out the 
unique treatment of the incident wave that one has to employ.

\begin{center}
{\large Acknowledgements}
\end{center}

We thank Seiji Sakoda for many valuable discussions on the AB 
scattering, in particular, for his observation that the unitarity 
relation holds with the Ruijsenaars scattering amplitude in the 
sense described in Sec. V. We acknowledge contributions made by 
Osamu Takeuchi in the early stage of this work. 
One of us (H.M.) is supported in part by Grant-in-Aid for Scientific 
Research \#09045036 under International Scientific Research Program, 
Inter-University Cooperative Research. He would like to thank Charlie 
Sommerfield for critical discussions on the Aharonov-Bohm scattering. 
He expresses gratitude to Center for Theoretical Physics, Yale 
University for hospitalities extended to him during his visits, 
which was partially supported also by Agreement between Tokyo 
Metropolitan University and Yale University on Exchange of Scholars 
and Collaborations, engaged in May 1996.

\begin{center}
{\large Appendix}
\end{center}

We derive the unitarity relation of the Aharonov-Bohm scattering
amplitude by following the method of Landau and Lifshitz \cite {LL}.
We consider the asymptotic form of the wave function
\begin{equation}
\psi=e^{-ikr\cos(\theta^\prime-\theta)-i\alpha(\theta^\prime-\theta)} +
\displaystyle\frac{e^{ikr}}{\sqrt{r}}f(\theta^\prime-\theta),
\label{eq:ablimit}
\end{equation}
which describes the scattering of the incident wave coming from the
direction of angle $\theta$ to the direction of angle $\theta^\prime$.
The angles $\theta$ and $\theta^\prime$ are measured from the $x$-axis.
The basic strategy of Landau and Lifshitz is to consider the
superpositions of the wave function (\ref {eq:ablimit}) with 
arbitrary weight functions $F(\theta)$, which also describe
certain scattering processes, and to demand the conservation
of the fluxes of the incoming and the outgoing waves.

To this goal we decompose the incident wave into the incoming and the
outgoing waves. Using the representation of the plane wave by the
sum of the Bessel functions
\begin{equation}
e^{ikr\cos{\theta}}=
\displaystyle\sum_{m = -\infty}^{\infty} i^me^{im\theta}J_m(kr),
\end{equation}
and noting the asymptotic form of the Bessel function
\begin{equation}
J_{\nu}(z)\; \displaystyle\yazi_{|z|\to\infty}\;
\sqrt{\frac{2}{\pi z}} \cos \left(z - \frac {(2 \nu + 1)\pi}{4} \right).
\end{equation}
one can show that
\begin{equation}
e^{-ikr\cos(\theta^\prime-\theta)-i\alpha(\theta^\prime-\theta)}
\displaystyle\ya
e^{-i\alpha(\theta^\prime-\theta)}
\left(\frac{2\pi}{kr} \right)^\frac{1}{2}
\left[
e^{i\left(kr-\frac{\pi}{4} \right)}\delta(\theta^\prime-\theta-\pi) +
e^{-i\left(kr-\frac{\pi}{4} \right)}\delta(\theta^\prime-\theta)
\right]
\end{equation}
We then obtain

\begin{equation}
\displaystyle\int F(\theta)\psi d\theta
\displaystyle\ya
\left(\frac{2\pi}{kr} \right)^\frac{1}{2}
\left[
e^{-i(kr-\frac{\pi}{4})}F(\theta^\prime) +
e^{i(kr-\frac{\pi}{4}-\alpha\pi)} \hat{S}F(\theta^\prime - \pi)
\right]
\label{eq:hadouzen}
\end{equation}
where $\hat{S}$ is the S-matrix;
\begin{equation}
\hat{S}=1+e^{i\frac{\pi}{4}}\sqrt{k}\hat{f},
\label{eq:smat}
\end{equation}
and the operator $\hat{f}$ acts as
\begin{equation}
\hat{f}F(\theta^\prime - \pi) =
\displaystyle\frac{e^{i\alpha\pi}}{\sqrt{2\pi}}
\int_{\theta^\prime-\pi}^{\theta^\prime + \pi} 
F(\theta)f(\theta^\prime-\theta)d\theta
\label{eq:fnoteigi}
\end{equation}

The unitarity of the S matrix, $\hat{S}^\dagger\hat{S}$ =1, which 
follows from the conservation of the probability current, leads to 
\begin{equation}
e^{-i\frac{\pi}{4}}\hat{f}-e^{i\frac{\pi}{4}}\hat{f}^\dagger =
i\sqrt{k}\hat{f}^\dagger\hat{f},
\end{equation}
By operating the both sides of this equation to 
$F(\theta^\prime-\pi)$ we obtain

\begin{eqnarray}
&e^{i\left(\alpha-\frac{\pi}{4} \right)}f(\theta^\prime-\theta) -
e^{-i\left(\alpha-\frac{\pi}{4} \right)}f^*(\theta-\theta^\prime)
\hspace{70mm}& \nonumber \\
& = i\displaystyle\sqrt{\frac{k}{2\pi}}
\left\{\int_{\theta^\prime-\pi}^{\theta}
f^*(\theta^{\prime\prime} - \theta^{\prime})
f(\theta^{\prime\prime}-\theta+\pi)d\theta^{\prime\prime} +
\int_{\theta}^{\theta^\prime+\pi}
f^*(\theta^{\prime\prime}-\theta^{\prime})
f(\theta^{\prime\prime}-\theta-\pi)d\theta^{\prime\prime} \right\}&
\end{eqnarray}
We set $\theta=0$ and replace $\theta^\prime$ by $\theta$ to derive 
the unitarity relation (\ref{eq:scattering3})


\end{document}